\documentclass[preprint]{elsarticle}

\usepackage{natbib}
\usepackage{lineno,hyperref}
\usepackage{siunitx}
\usepackage{verbatim}
\usepackage{graphicx}
\usepackage{caption}
\usepackage{subcaption}
\usepackage{amssymb, amsmath}
\usepackage{setspace}
\modulolinenumbers[5]

\journal{Physics of the Dark Universe}

\begin{document}

\begin{frontmatter}
\title{First background-free limit from a directional dark matter experiment: results from a fully fiducialised DRIFT detector}

\author[wellesley]{J.~B.~R.~Battat}
\author[csu]{J.~Brack}
\author[shef]{E.~Daw}
\author[csu]{A.~Dorofeev}
\author[shef]{A.~C.~Ezeribe}
\author[oxy]{J.-L.~Gauvreau}
\author[unm]{M.~Gold}
\author[csu]{J.~L.~Harton}
\author[oxy,unm]{J.M.~Landers}
\author[oxy]{E.~Law}
\author[unm]{E.~R.~Lee}
\author[unm]{D.~Loomba}
\author[oxy]{A.~Lumnah}
\author[unm]{J.~A.~J. Matthews}
\author[unm]{E.~H.~Miller}
\author[oxy]{A.~Monte}
\author[shef]{F.~Mouton}
\author[edinburgh]{A.~StJ.~Murphy}
\author[boulby]{S.~M.~Paling}
\author[unm]{N.~Phan}
\author[shef]{M.~Robinson}
\author[shef]{S.~W.~Sadler\corref{mycorrespondingauthor}}
\author[shef]{A.~Scarff}
\author[csu]{F.~G.~Schuckman II}
\author[oxy]{D.~P.~Snowden-Ifft}
\author[shef]{N.~J.~C.~Spooner}
\author[shef]{S.~Telfer}
\author[hawaii]{S.~E.~Vahsen}
\author[shef]{D.~Walker}
\author[csu]{D.~Warner}
\author[shef]{L.~Yuriev}
\cortext[mycorrespondingauthor]{Corresponding author}
%%%\ead{support@elsevier.com}

\address[wellesley]{Department of Physics, Wellesley College, 106 Central Street, Wellesley, MA 02481, USA}
\address[csu]{Department of Physics, Colorado State University, Fort Collins, CO 80523-1875 USA}
\address[shef]{Department of Physics and Astronomy, University of Sheffield, S3 7RH, UK}
\address[oxy]{Department of Physics, Occidental College, Los Angeles, CA 90041, USA}
\address[unm]{Department of Physics and Astronomy, University of New Mexico, NM 87131, USA}
\address[edinburgh]{School of Physics and Astronomy, University of Edinburgh, EH9 3FD, UK}
\address[boulby]{STFC Boulby Underground Science Facility, Boulby Mine, Cleveland, TS13 4UZ, UK}
\address[hawaii]{Department of Physics and Astronomy, University of Hawaii, Honolulu, HI 96822, USA}

\begin{abstract}
The addition of O$_2$ to gas mixtures in time projection chambers containing CS$_2$ has recently been shown to produce multiple negative ions that travel at slightly different velocities. This allows a measurement of the absolute position of ionising events in the $z$~(drift)~direction. In this work, we apply the $z$-fiducialisation technique to a directional dark matter search. We present results from a 46.3 live-day source-free exposure of the DRIFT-IId detector run in this new mode. With full-volume fiducialisation, we have achieved the first background-free operation of a directional detector. The resulting exclusion curve for spin-dependent WIMP-proton interactions reaches $1.1$~pb at $100$~GeV/c$^2$, a factor of 2 better than our previous work. We describe the automated analysis used here, and argue that detector upgrades, implemented after the acquisition of these data, will bring an additional factor of $\gtrsim 3$ improvement in the near future.
\end{abstract}


\end{frontmatter}
\newpage 
%\linenumbers

\section{Introduction}
\label{sec:intro}
There is strong evidence from a variety of sources to suggest that 85\% of the Universe's matter is in the form of dark matter (DM)~\cite{Beringer2012}. One possibility favoured by theories beyond the Standard Model of particle physics is that DM consists of Weakly Interacting Massive Particles (WIMPs)~\cite{Feng2010}.  As such, a large, international effort has been underway for decades to search for the rare, low-energy recoil events produced by WIMP interactions \cite{Beringer2012}.  Several groups have published results in which a handful of events appear above calculated background \cite{Agnese2013,Angloher2012}.  Meanwhile, two groups have results consistent with the detection of WIMPs through the use of the annual modulation signature \cite{Bernabei2010, Aalseth2014}. These latter results are inconsistent with other limits \cite{Akerib,Aprile2012,PhysRevLett.112.241302} under nearly all possible scenarios.

The goal of directional dark matter detectors is to provide a `smoking gun' signature of DM \cite{Ahlen2010}. Such experiments seek to reconstruct not only the energy, but also the direction of WIMP-induced nuclear recoils, thereby confirming their signals as galactic in origin. Numerous studies have shown the power of a directional signal \cite{Copi1999,Morgan2005}.  Instead of order $10^4$ events required for confirmation via the annual modulation signature, only of order $10 - 100$ events are required with a directional signature \cite{Green2007}. Additionally, instead of easily mimicked annual modulation, the directional signal is fixed to the galactic coordinate system and is therefore immune to false-positive detections. In recent years, several ideas for directional detection have been proposed and revived~\cite{Nygren2013,Belli1992,Cappella2013,Spooner1997,Shimizu2003}. At present, however, the only demonstrated directional technology is recoil tracking in gaseous time projection chambers \cite{Ahlen2010}.

The Directional Recoil Identification From Tracks (DRIFT) collaboration pioneered the use of low-pressure gas TPCs to search for this directional signal~\cite{Snowden-Ifft2000}. Uniquely, DRIFT utilises the drift of negative ions (in particular, CS$_2^-$) to transport the ionisation to the readout planes with diffusion at the thermal limit~\cite{Snowden-Ifft2013}. DRIFT has systematically progressed through several epochs of focused R\&D: (1) demonstration of stable operation of the first negative ion TPC (NITPC) and, at the \si{\metre \cubed} scale, the largest directional detector~\cite{Snowden-Ifft2000} (2) proof-of-principle of directionality~\cite{Burgos2009,Burgos2009a} and (3) identification and elimination of detector backgrounds~\cite{burgos2007,Battat2014,Brack2014,Brack2014a}. Presented here are results that constitute a major step forward in stage (3): the first full-volume fiducialiation and background-free operation of a directional dark matter detector.

\begin{figure}
\centering
\begin{subfigure}{.42\textwidth}
\centering
\includegraphics[width=0.95\textwidth]{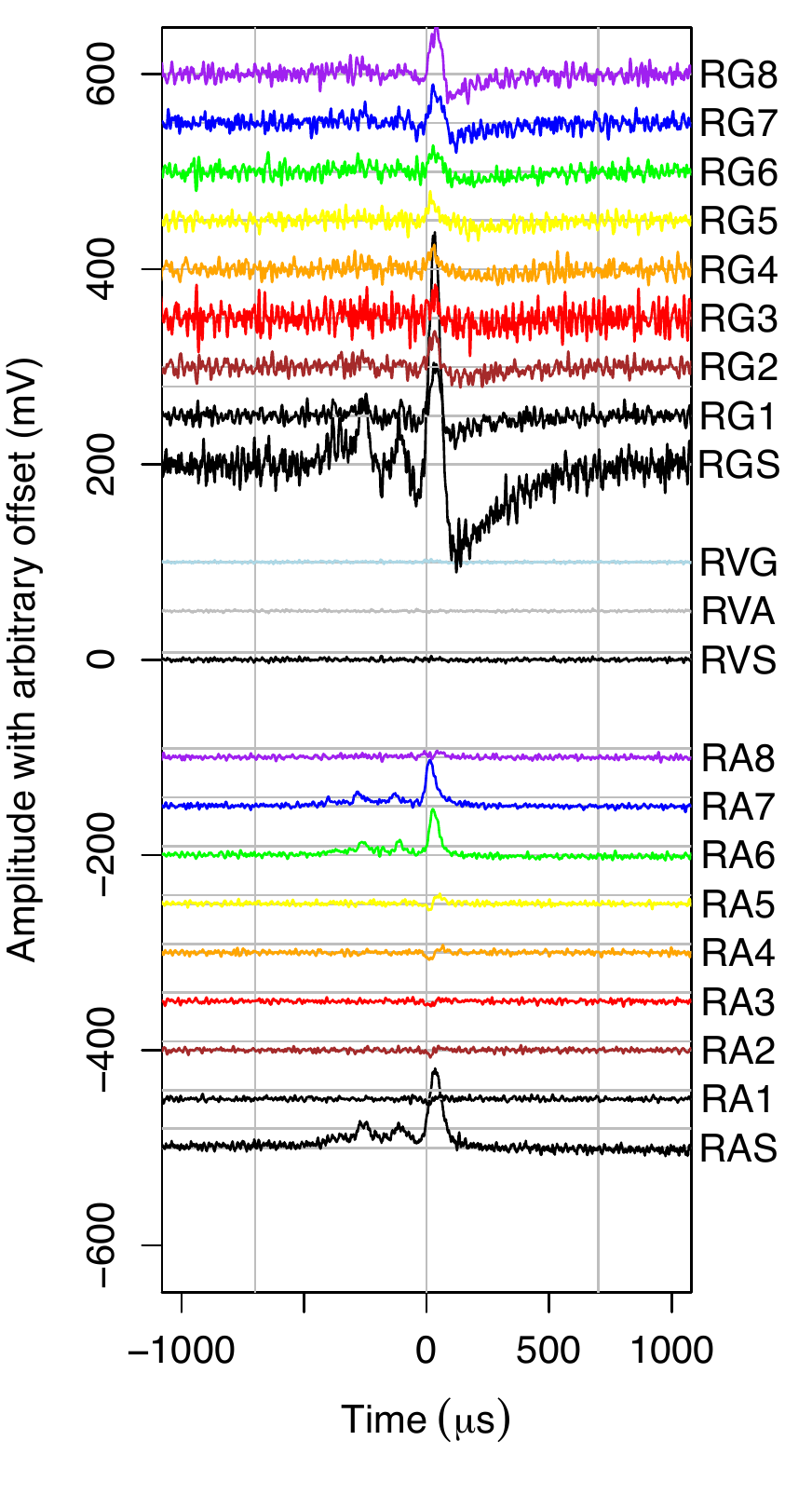}
\caption{}
\label{fig:mincar1}
\end{subfigure}%
\begin{subfigure}{.58\textwidth}
\centering
\includegraphics[width=0.95\textwidth]{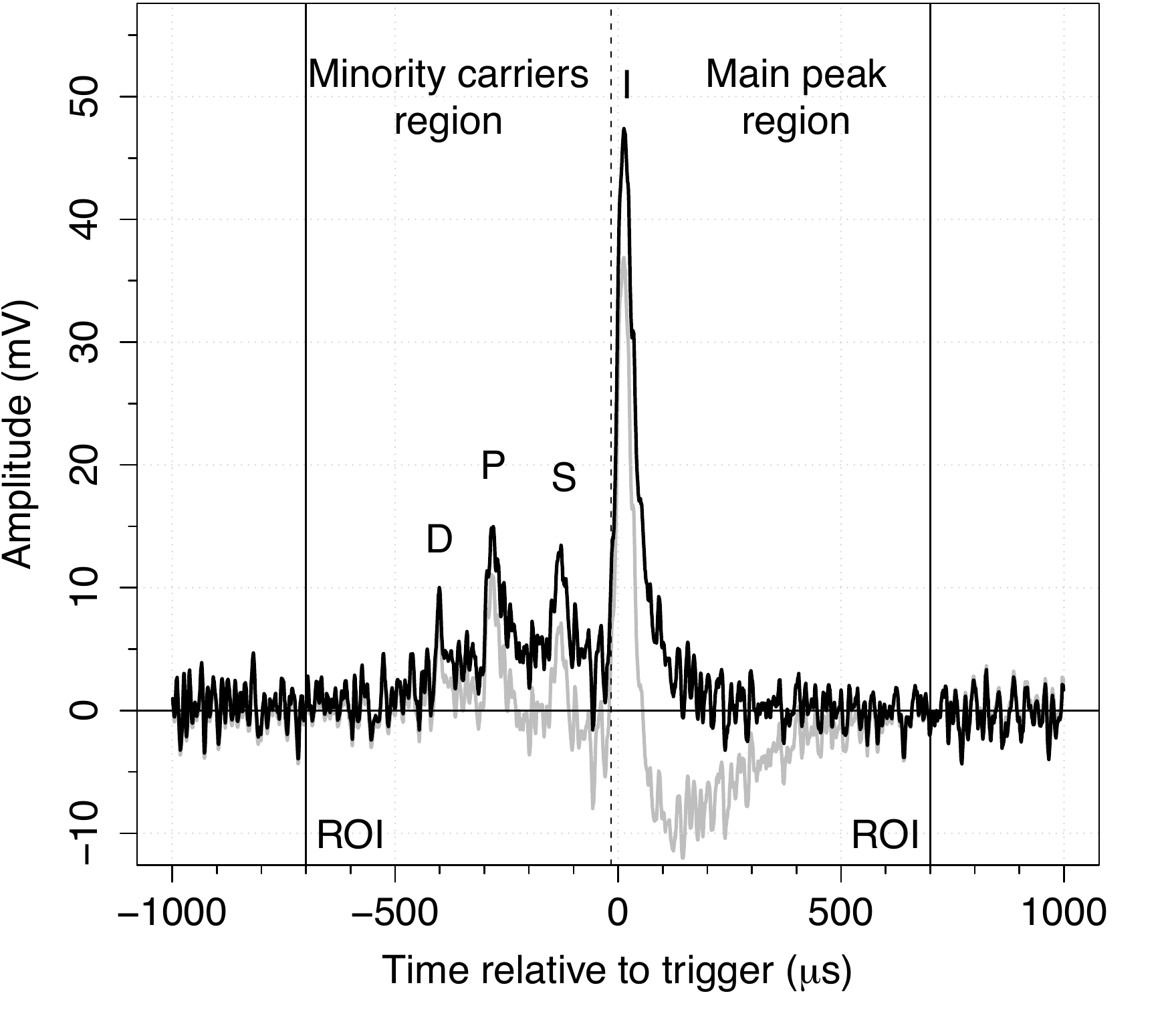}
\caption{}
\label{fig:mincar2}
\end{subfigure}
\caption{a) Example detector output after pulse shaping for a typical neutron recoil candidate event, demonstrating the minority carrier signature. The event produced $4980$~NIPs at a distance of $35.6\pm0.2$~cm from the right MWPC during a neutron-calibration run. In the waveform labels, `R' stands for `right', `A' for `anode', `G' for `grid', `V' for `veto' and `S' for `sum'. Horizontal grey lines above the coloured anode waveforms indicate the analysis threshold. b) labeled detail of the channel RA7 (blue) waveform before (grey) and after (black) application of the undershoot removal algorithm discussed in the text. The dotted vertical line separates the minority (S, P, D) peaks from the main (I) peak.}
\label{fig:mincar}
\end{figure}

Background-free operation was made possible by the addition of 1 Torr of O$_2$ to the nominal 30:10~Torr CS$_2$:CF$_4$ DRIFT gas. This produces several species of so-called `minority carriers' that drift with slightly different velocities relative to the single species observed with the regular gas mixture~ \cite{Snowden-Ifft2013c}. An example is shown in Figure~\ref{fig:mincar}. With the arrival time difference between the peaks proportional to the distance from the readout plane, these minority carriers enable a measurement of the distance, $z$, from the readout plane to the ionising event. This $z$-coordinate measurement enables the removal of all nuclear recoil backgrounds due to radioactive decays on either the central cathode (high-$z$) or readout planes (low-$z$). The detector fiducialisation preserves a large nuclear recoil efficiency, thereby expanding our signal window relative to our previously published limit~\cite{Daw2012}.

\section{DRIFT-IId detector and science runs}
\label{sec:drift}
The DRIFT experiment is sited at a depth of 1.1~km in the STFC Boulby Underground Science Facility~\cite{Paling2012}, which provides 2805~m.w.e. shielding against cosmic rays. The TPC is housed inside a stainless steel cubic vacuum vessel, surrounded on all sides with 44~\si{\gram \per \centi \metre \squared} of polypropylene pellets to shield against neutrons from the cavern walls. The vessel was filled with a mixture of 30:10:1~Torr CS$_2$:CF$_4$:O$_2$ gas, and sealed for the duration of each run. This departure from the normal mode of operation, in which gas is flowed at a constant rate of one complete vacuum vessel change ($590$~g) \si{ \per \day}, was necessary due to safety concerns over sources of ignition in the constant flow system. These concerns have since been addressed with modifications to the gas system.

The DRIFT-IId NITPC consists of a thin-film (0.9~\si{\micro \metre} aluminised Mylar), texturised central cathode \cite{Brack2014} at a potential of -31.9~kV faced on either side by two 1~\si{\metre \squared} multi-wire proportional chambers (hereafter, the `left' and `right' MWPCs) at a distance of 50~cm. In this way, two 50-cm-long drift regions are defined. A field cage of 31 stainless steel rings on either side steps down the voltage smoothly between the central cathode and the MWPCs to ensure a uniform electric field of 580~\si{\volt \per \centi \metre} throughout the drift regions. The MWPCs are made up of a central grounded anode plane of 20~\si{\micro \metre} diameter stainless steel wires with 2~mm pitch, sandwiched between two perpendicular grid planes of 100~\si{\micro \metre} wires at -2884~V, again with 2~mm pitch and separated by 1~cm from the anode plane. A full description of the detector can be found in Ref.~\cite{Alner2005}.

Both the inner grid and anode planes have every eighth wire joined together and read out as one, such that a single `octave' of wires reads out $8 \times 2 = 16$~mm in $x$  and $y$: large enough to contain the recoil events of interest. The outermost 52 (41) wires of the 512 total on the inner grid (anode) planes are grouped together into $x$ ($y$) veto regions, reducing the fiducial volume of the detector to 0.80~\si{\metre \cubed}. The anode and grid veto signals are summed to produce a `veto sum' waveform. All signals are pre-amplified inside the vacuum vessel by Cremat CR-111 preamplifiers, then shaped with a time constant of 4~\si{\micro \second} and amplified by Cremat CR-200 Gaussian shaping amplifiers. Finally, the signals pass through a high-pass filter with time constant 110~\si{\micro \second}, and are digitised by a 14-bit National Instruments PXI-6133 ADC with input voltage range $-1. 25 < V_{\text{in}} < 1.25$~V and sampling rate of 1~MHz, chosen to give good energy and time resolution for signals in the energy range, and of the spatial extent, of interest.

The WIMP-search dataset consists of 46.3~live-days of sealed, fully shielded, source-free operation, with 33.2~g of fluorine in the $0.8$~m$^3$ active volume as a target for spin-dependent (SD) WIMP interactions. The ionisation measurement was calibrated automatically every 6 hours using 5.9~keV X-rays from two shuttered $^{55}$Fe sources mounted behind each MWPC. Additionally, 3.2~live-days of neutron-calibration data were interspersed with the science runs. During the neutron calibrations, a $\sim1500$~Bq $^{252}$Cf neutron source was placed on top of the vacuum vessel and within the shielding, generating several thousand nuclear recoil events used to determine the WIMP detection efficiency~\cite{Burgos2009}. In addition to providing stability information, these data were used to validate the minority peak analysis.

\section{Data analysis}
\label{sec:data}
During the 46.3~live-day WIMP-search dataset, the detector operated in the normal trigger mode: if any of the voltages on the 16 anode channels (8 in the right MWPC and 8 in the left MWPC) exceeded a fixed threshold of 30~mV, then the signals from all anode, grid and veto channels (3~ms pre-trigger and 10~ms post-trigger) were written to disk as an `event.' The typical rate of events was $\sim 1$~Hz. Because the total ionisation produced is shared approximately equally between the main and minority peaks, and because the majority of triggered events are near threshold on the main peak, the energy threshold in this run was approximately a factor of 2 larger than in our previously published limit~\cite{Daw2012}, run with the same voltage threshold. An updated triggering scheme has since resolved this issue.

A temporal region of interest (ROI) was defined between $- 700$~\si{\micro \second} and $+700$~\si{\micro \second} relative to the trigger time. A basic cut was applied to remove events that saturated the ADCs, and the waveforms were subject to noise reduction algorithms that a) remove 55~kHz noise caused by the high voltage power supply, by applying a notch filter to the frequency spectrum, and b) reduce baseline wander by subtracting a sine function with a fixed frequency of 50\,Hz from the waveform. These noise reduction techniques allow for an analysis threshold at $9$~mV: significantly below the 30~mV hardware threshold.\footnote{The 30\,mV threshold triggers the event, typically on the I peak, while the 9\,mV threshold then allows the identification of the D, P and S peaks.} The shaping electronics described in Section~\ref{sec:drift} cause the signals to undershoot the baseline, mainly due to AC coupling. This undershoot was corrected prior to event reconstruction using a two-stage undershoot removal algorithm implemented in software. The algorithm uses time constants measured on a channel-by-channel basis. Figure~\ref{fig:mincar2} shows that this filter successfully restores the baseline. Further basic cuts were applied to remove events that a) crossed a 15~mV veto sum threshold ($x$ and $y$ fiducialisation), b) extended outside the temporal ROI, c) had hits on both sides of the detector, d) had non-contiguous channel hits, e) had hits on all 8 anode channels (implying a range of 16~mm: far longer than the several-mm tracks expected from DM-nucleus interactions), and f) had risetime $<3$~\si{\micro \second}, consistent with an impulse charge deposition. These basic cuts are referred to collectively as `Stage 1' cuts, and are designed to have very high acceptance for nuclear recoils.

To reconstruct the number of ion pairs (NIPs) in an event, we select the anode channels that cross the analysis threshold of 9~mV, and integrate the anode waveform across the ROI. The final NIPs value for the event is the sum of these integrated values, scaled by a calibration constant. This constant was calculated every 6 hours using the 5.9~keV X-rays from the $^{55}$Fe calibration sources \cite{burgos2007} and a W-value of $25.2$~eV. The W-value is the average energy required to produce an electron-ion pair in the gas. The W-value for a 30:10~Torr CS$_2$:CF$_4$ mixture was used for this calculation \cite{Pushkin2009} because no measurement has yet been made for 30:10:1~Torr CS$_2$:CF$_4$:O$_2$. This will be addressed in a forthcoming paper. However, the results of Refs.~\cite{Haeberli1953,Moe1957,Bronic1991} suggest that the addition of 1~Torr O$_2$ will change the W-value by $\sim 1\%$, which is a factor of two smaller than the uncertainty of the W-value used here~\cite{Pushkin2009}. Fluctuations in the $^{55}$Fe energy measurement were observed at the $\sim 4\%$ level over the course of the full 46.3~live-day dataset. Table~\ref{tab:nips} shows how the NIPs yield varies with energy of the fluorine recoil. It is based on calculations in Ref.~\cite{Hitachi2008}, as validated experimentally in~\cite{Daw2012}.
\begin{table}[t]
\begin{center}
\resizebox{6.5cm}{!} {
\begin{tabular}{|c|c|} \hline
\textbf{Fluorine recoil energy (keV$_{\text{r}}$)} 	&	\textbf{NIPs} 	\\ \hline
10							&		140		\\ 
20							&		332		\\ 	
50							&		1055		\\ 
100							&		2528		\\ 
150							&		4165		\\ 	
200							&		5852		\\ \hline 
\end{tabular}
}
\caption{NIPs yield as a function of energy for fluorine recoils, from Ref.~\cite{Hitachi2008}.}
\label{tab:nips}
\end{center}
\end{table}

The use of $z$-fiducialisation has increased the signal acceptance relative to Ref.~\cite{Daw2012} by replacing a set of low-efficiency cuts that had been necessary to remove background events originating from the electrodes, with a reduced set of straightforward, high-efficiency cuts described below. The ratio between the ionisation measured in the minority peaks to that measured in the I peak on the channel with the highest maximum voltage in the ROI (see Figure~\ref{fig:mincar}) was found to be a powerful parameter for discriminating between nuclear recoils caused by calibration neutrons and background events such as sparks~\cite{Snowden-Ifft2013c}. Events for which this ratio was $< 0.4$ were cut. One further high-efficiency ($\sim 97\%$) cut was added to ensure that the ionisation detected on the grid was in agreement with that detected on the anode, which removed a residual population of oscillatory background events described in Ref.~\cite{Snowdenifft2003}.

An event that passed the preceding set of cuts had its maximum-amplitude channel's waveform passed to a peak-fitting algorithm, which used a three-Gaussian fit to find the arrival times of the I, S and P peaks at the anode. The D peak was not used since its amplitude was often within the noise. The time difference between any two peaks can be used to calculate the $z$ position of the event from:
\begin{equation}
z  = (t_a - t_b  ) \frac{v_a v_b}{( v_b- v_a )} \text{.}
\label{eqn:z}
\end{equation}
Here, $a$ and $b$ represent two different carrier peaks (I, S or P), and $t$ and $v$ are the arrival time relative to trigger, and drift velocity, respectively. An event passed the $z$-reconstruction algorithm if three equidistant peaks were found.  Of those peaks, the earliest and latest in time were assumed to be P and I, respectively. In practise, the P and I peaks were used to calculate the
$z$ position of the event (via Equation~\ref{eqn:z}), because the S peak is suppressed for high-$z$ events \cite{Snowden-Ifft2013c}. Events were cut if three equidistant peaks were not found, or if the fitting procedure produced unphysical results.  This removed all events from the WIMP-search run other than a small population of well-understood radon progeny recoils (RPRs) and low-energy alphas (LEAs) originating near the cathode \cite{burgos2007}, and preserved $\sim70\%$ of recoil-like events from the neutron-calibration runs. The $30 \%$ that were lost originated either at high $z$ (the S peak is suppressed, so three equidistant peaks were not found), or at small $z$ (our current automated analysis could not identify three separate peaks because of the temporal overap of I, S, and P).

\begin{figure}
\begin{center}
\includegraphics[width=\textwidth]{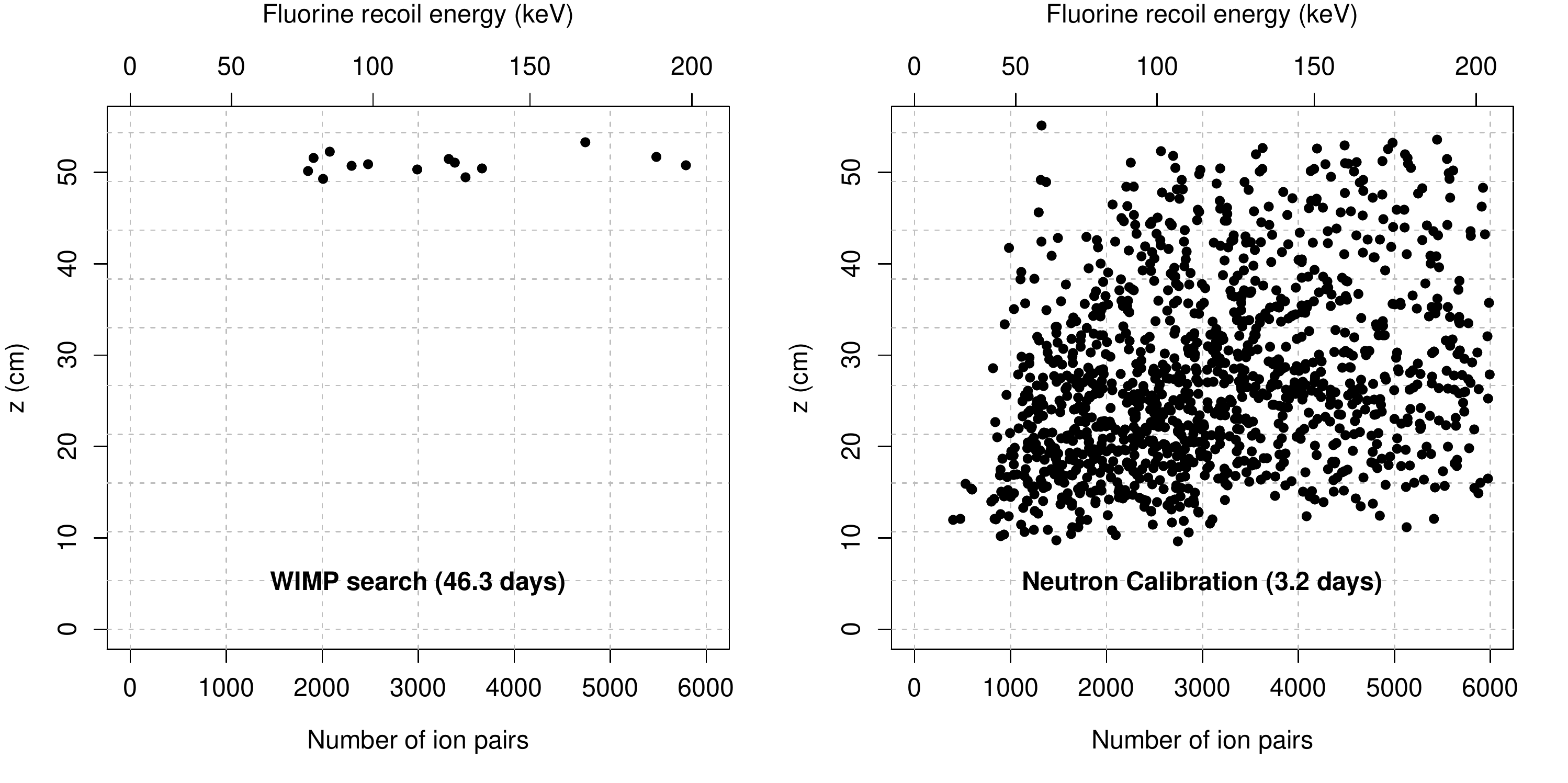}
\caption{$z$ vs. NIPs distributions of events in the WIMP search and calibration neutron datasets (with top axis giving nuclear recoil energy inferred from Ref.~\cite{Hitachi2008}). Events recorded during WIMP-search operation (left panel). No background events appear at any energy in the bulk fiducial volume below $z = 48.4$\,cm. Neutron-calibration data (right panel) shows that DRIFT is sensitive to neutron-induced recoils, and hence WIMP interactions, in this region. The grid lines are drawn to match the bins in the efficiency map of Figure~\ref{fig:map}. The absence of events below $z \sim 10$\,cm and the $z$-dependence of the low-energy cutoff in the neutron-calibration data are described in the text.}
\label{fig:znips}
\end{center}
\end{figure}

The relative velocities $v_{a,b}$ used in the $z$ measurement were calibrated in-situ during the science runs using a selected sample of RPR/LEA events.  From previous work \cite{burgos2007}, these events are known to originate on the central cathode at $z = 50$~cm.  This calibration was needed because of contamination of the fill gas in the sealed mode of operation.
Based upon the uncertainty on the I and P peak means in the three-Gaussian fit, the precision in the measurement of $z$ is 1.6~mm.  The reconstructed $z$ positions have a spread of $\pm 2$\,cm.
Spatial variations in the height of the thin-film cathode (ripples) \cite{driftThinFilm} cause part of this spread, and further studies are underway to explain the remaining scatter.  For the purposes of this work, we restrict our fiducial volume to $z<48$\,cm.  The $z$ vs. NIPs distributions for events in the WIMP-search (46.3~days) and neutron-calibration (3.2~days) datasets are shown in Figure~\ref{fig:znips}.

\section{Nuclear recoil efficiency}
\label{sec:map}
As discussed in Ref.~\cite{Burgos2009}, elastic recoils from $^{252}$Cf neutrons provide a useful calibration data set. A total of 3.2 live-days of neutron-calibration data were taken in 5 dedicated runs interspersed with WIMP-search operation, during which time the $^{252}$Cf source was placed on top of the vacuum vessel, inside the shielding. The activity of the source at the time of the exposures varied from 1500 to 1410 neutrons/s with a systematic uncertainty of about 80 neutrons/s~\cite{burgos2007}. To inhibit gamma ray interactions, the source was contained within a cylindrical lead canister of wall thickness 1.3~cm.

\begin{figure}
\begin{center}
\includegraphics[width=0.65\textwidth]{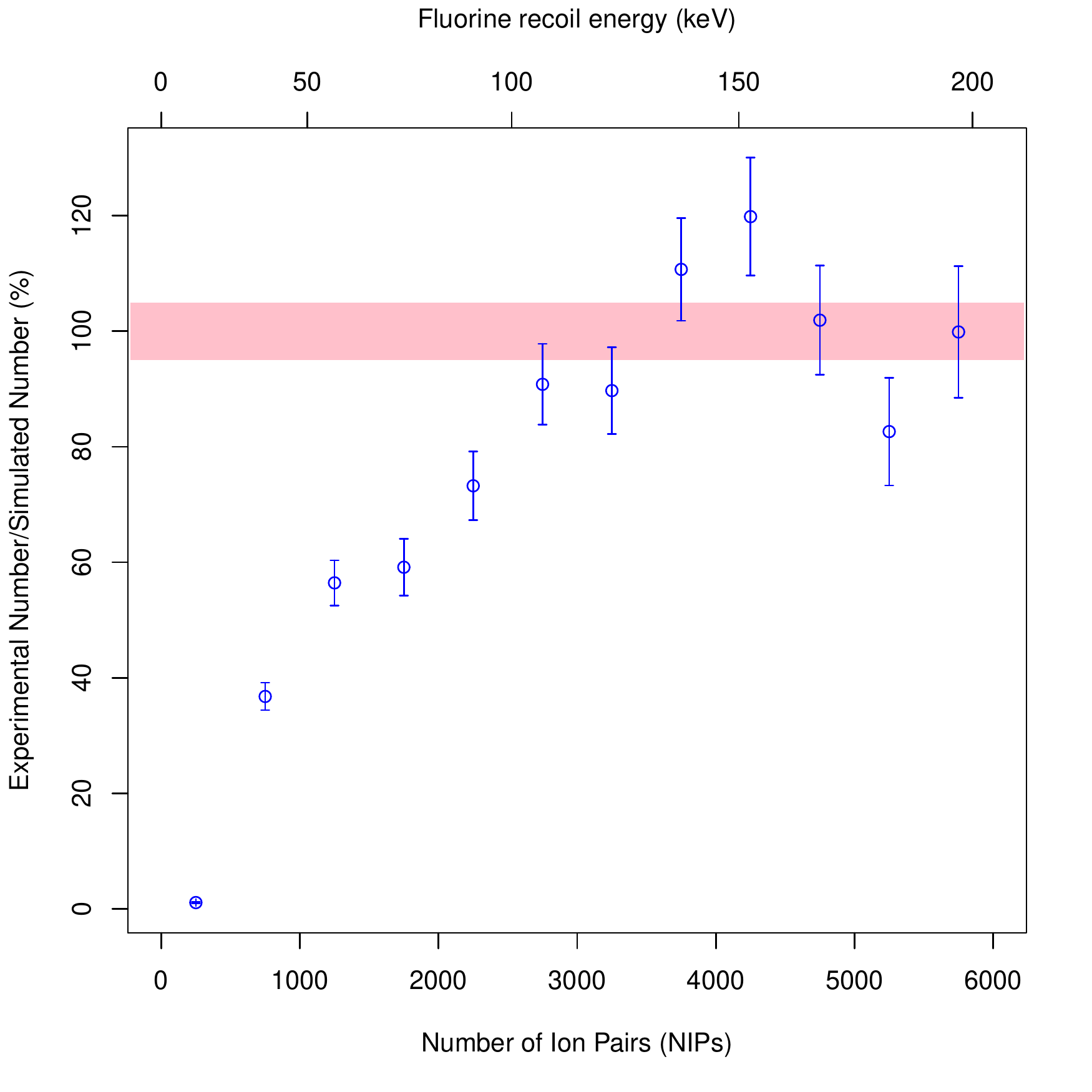}
\caption{The ratio of the number of background-subtracted events passing the Stage~1 cuts as a function of NIPs to a scaled number of simulated events, with 1$\sigma$ error bars.  The scaling factor was determined from the data between 3000 and 6000 NIPs. The horizontal pink band represents the systematic uncertainty of the $^{252}$Cf source strength.
  The efficiency reaches 100\% at a higher energy than in Ref.~\cite{burgos2007} due to the higher trigger threshold discussed in the text.  The top axis gives nuclear recoil energy as inferred from Ref.~\cite{Hitachi2008}.}
\label{fig:xxx}
\end{center}
\end{figure}

A detailed GEANT4 simulation of the Cf-252 exposure was used to obtain the distribution of expected neutron recoils in the $z$ vs. NIPs parameter space.  The simulation included the shielding, the location of the source, the Pb around the source and all detector components. Elastic and inelastic interactions with nuclei within the fiducial volume had their recoil types, initial energies, and initial interaction locations recorded. The initial recoil energies were converted to NIPs using quenching factors calculated by Hitachi~\cite{Hitachi2008} and validated experimentally in~\cite{Daw2012}. A similar simulation of the DRIFT-IIa detector showed $100 \pm 2\%$~(statistical)~$\pm 5\%$~(systematic) efficiency after the Stage~1 cuts and background subtraction for large NIP events~\cite{burgos2007}. Here, the total number of neutrons emitted was unfortunately not recorded, and so we took advantage of the previous result to normalise the simulation to the data. Stage~1 cuts were applied to the data and the number of background-subtracted events in 500-NIP bins from 0 to 6000~NIPs were calculated.
The simulated data was normalized so that the ratio of the expected and measured number of events in the 3000--6000 NIPs region after Stage 1 cuts was unity, as shown in Figure~\ref{fig:xxx}.

\begin{figure}
\begin{center}
\includegraphics[width=0.72\textwidth]{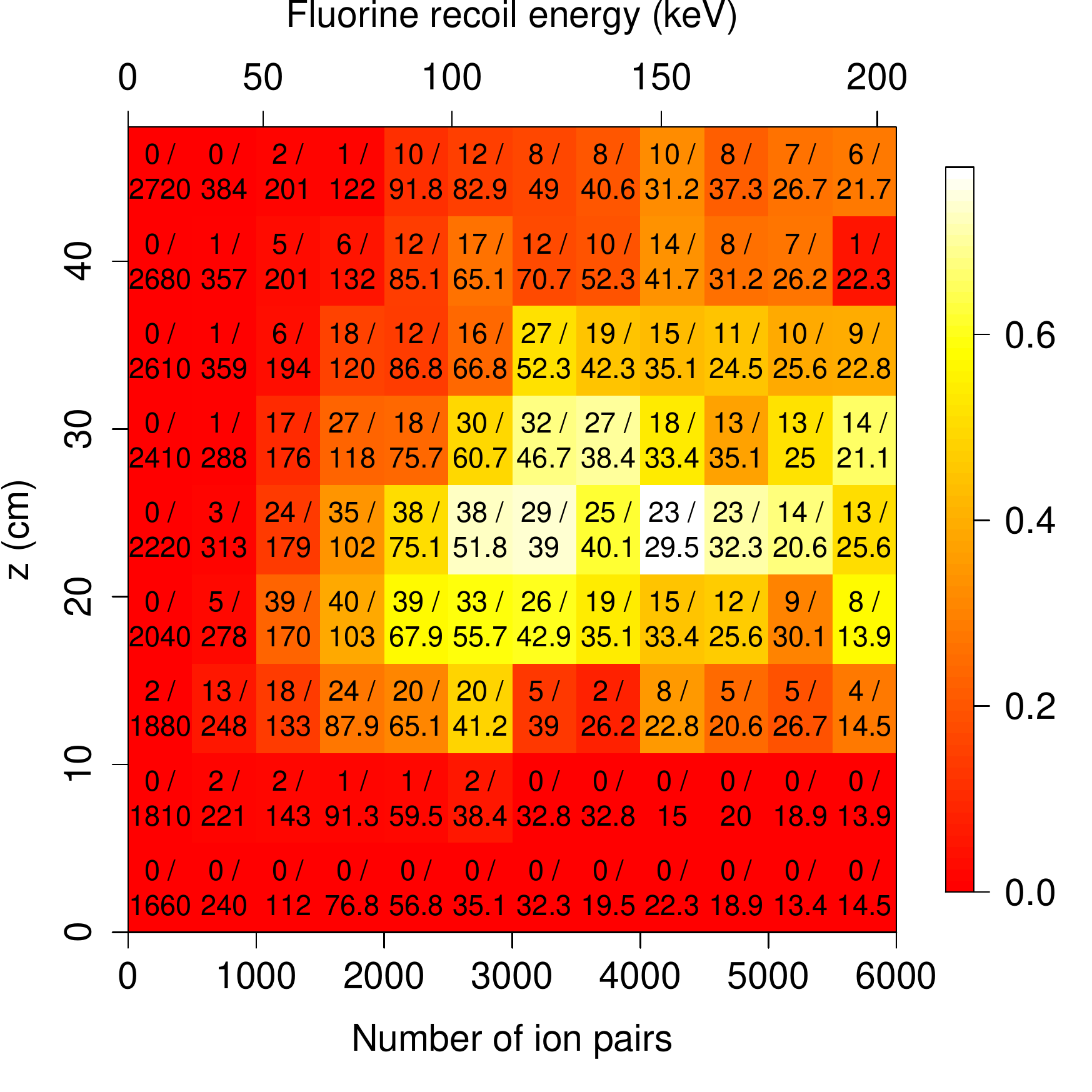}
\caption{An efficiency map for recoil detection including all analysis cuts.  Each cell shows the number of detected recoils over the number expected from the GEANT4 simulation after application of the normalisation described in the text.  The top axis gives nuclear recoil energy as inferred from Ref.~\cite{Hitachi2008}.  The overall event rate is expected to be higher at large $z$ due to the placement of the $^{252}$Cf source near the central cathode.}
\label{fig:map}
\end{center}
\end{figure}

An efficiency map was then calculated as follows. The parameter space of $z$ vs. NIPs was subdivided into cells as shown in Figure~\ref{fig:znips}, and the number of accepted neutron recoils from the analysis, including all cuts, calculated for each cell. This number was divided by the scaled, simulated number of events for each cell to provide an efficiency map as shown in Figure~\ref{fig:map}. Note that by far the most important ratios for limit setting are those at the threshold of detection ($\sim 1000$~NIPs). The advantages of this efficiency map are ease of calculation, freedom from model-dependent parameters (other than GEANT4), complete inclusion of any biases from analysis including difficult-to-model biases from peak fitting, and transparency for others wishing to make use of DRIFT data.

The efficiency is zero below $z \sim 10$~cm for two reasons, both related to the temporal overlap of minority peaks at short drift distances. First, events without a valid $z$-coordinate reconstruction are cut from the analysis, and the $z$-reconstruction algorithm requires a detection of three equally spaced peaks (see Section~\ref{sec:data}).  Second, if the minority and I peaks overlap, then the minority carrier ionization cannot be reconstructed separately, and the event fails the cut on the ratio of the ionization in the minority peak to the ionization in the I peak, as discussed in Section~\ref{sec:data}.

The events at low NIPs are cut because of the hardware threshold.  The positive slope of the left-hand edge of the neutron events shown in Figure~\ref{fig:znips} is due to diffusion-broadening of the signal, which causes the peak to fall below threshold. An improved triggering scheme that monitors the integral of the signal, currently under development, could recover these events.

\section{WIMP recoil simulation and limits}
\label{sec:limit}
\begin{figure}
\begin{center}
\includegraphics[width=0.8\textwidth]{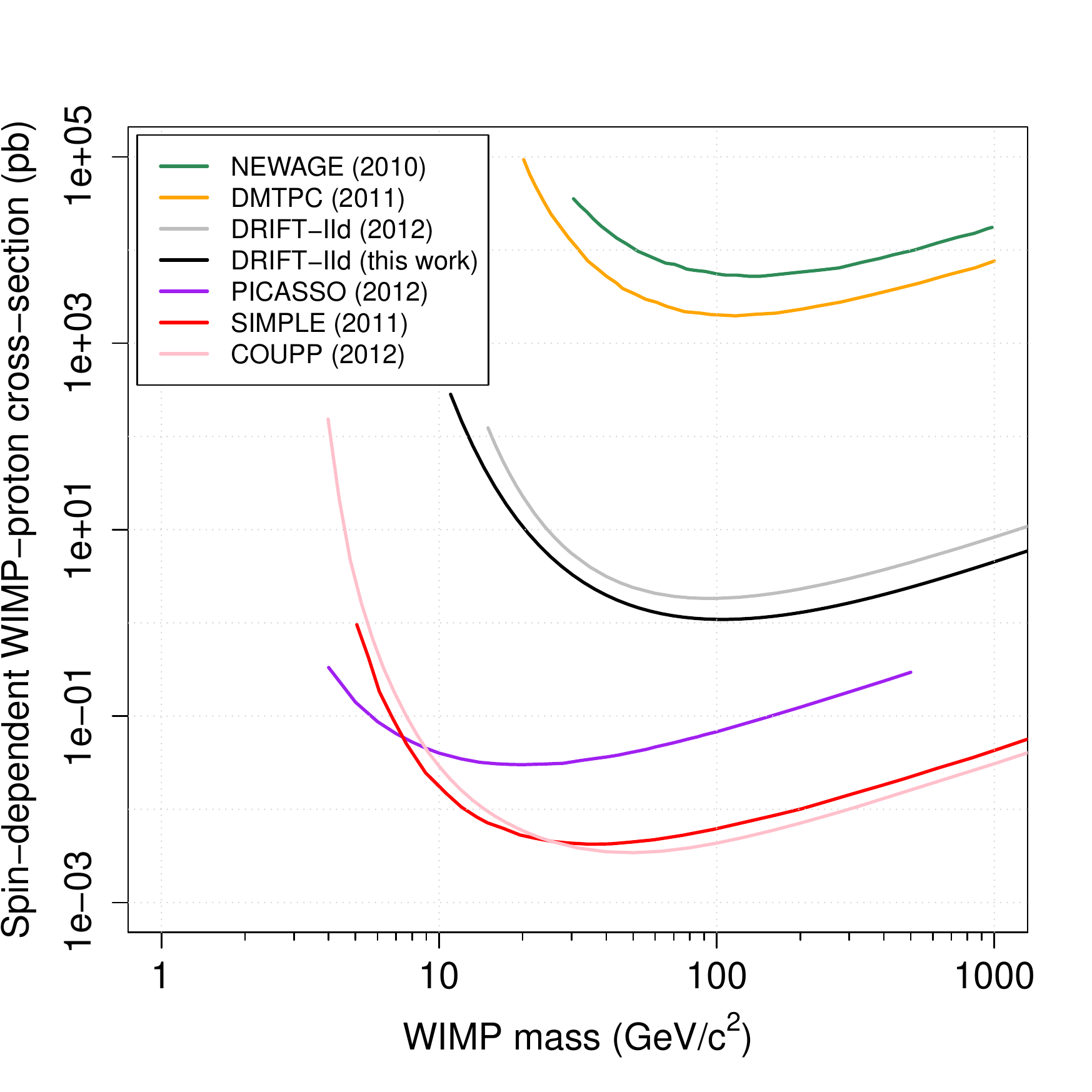}
\caption{Limits on the spin-dependent WIMP-proton interaction cross section. The result of this work is shown in black, and the previous limit curve from the DRIFT experiment using a similar live time is shown in grey~\cite{Daw2012}. All limits from other directional DM detectors (DMTPC~\cite{Ahlen2011} and NEWAGE~\cite{Miuchi2010}) are also shown, as are those from the leading non-directional detectors (PICASSO~\cite{Archambault2012}, SIMPLE~\cite{Felizardo2012} and COUPP~\cite{Behnke2012}).}
\label{fig:limits}
\end{center}
\end{figure}

The calculation of limits proceeded as follows.
For each of 32 WIMP masses, 9,000 fluorine recoils were simulated, using the parameters and equations found in Ref.~\cite{Lewin1996} ($v_0 = 230$~km/s, $v_E = 244$~km/s and $v_{\text{esc}} = 600$~km/s).  The recoils were distributed uniformly in $z$.  The number of fluorine recoils in each cell of the efficiency map were counted and then multiplied by the cell efficiencies.  The ratio of the resulting summed numbers divided by 9,000 then allowed us to calculate the overall detection efficiency for each WIMP mass.  The expected number of detected recoils for each WIMP mass was found by multiplying the nominal rate, calculated using the parameters and equations in Ref.~\cite{Lewin1996} (with $\rho_D = 0.3$~GeV/c$^2$/cm$^3$), by the exposure time (46.3~days) and by the detection efficiency for a fixed WIMP-nucleus cross-section.  This cross-section was then scaled to produce 2.3 detected fluorine recoils to provide a 90\% confidence limit cross-section.  Finally, this was converted to a WIMP-proton spin-dependent cross section using the method of Tovey et al. \cite{Tovey2000}.  The resulting data was then smoothed to produce the exclusion curve shown in Figure~\ref{fig:limits}.

\section{Results and discussion}
\label{sec:results}
The limit curve shown in Figure~\ref{fig:limits} improves upon the previous limit from DRIFT by a factor of two~\cite{Daw2012}, and was achieved with comparable live-time (46.3~days vs 47.4~days). The increase in efficiency driving this improvement is due to the removal of the low-efficiency cuts and the expansion of the signal region. 
However, because the fixed-amplitude trigger threshold used in this analysis was the same as in the previous work, the effective ionization threshold has increased. This is because $\sim \frac{1}{2}$ of the charge is transferred out of the I peak and into the minority carrier peaks, as discussed above. 
 
Subsequent test runs have demonstrated stable operation of DRIFT-IId with a threshold that is a factor of two lower than used in this analysis without triggering on electronic noise. This should improve the efficiency by a factor of $ \gtrsim 3$.  In addition, work is underway to trigger on the pulse integral (recoil energy), rather than on the pulse amplitude. Modifications to the gas flow system have been identified that will permit operation with continuous-flow of the CS$_2$:CF$_4$:O$_2$ mixture, rather than as a series of sealed runs. This continuous-flow mode of operation will ensure that the oxygen partial pressure, and therefore the relative minority carrier ionisation, remain stable. This is important for maximising the efficiency of the peak-finding algorithm. Finally, improvements to the automated peak-finding analysis are in development, which are expected to improve the efficiency further.

\section{Conclusion}
\label{sec:conclusion}
Background-free operation of a direction-sensitive WIMP detector with full 3D fiducialisation has been demonstrated for the first time. An analysis of 46.3 live-days of data taken with the directional dark matter detector {DRIFT-IId} with fiducialising O$_2$ additive has produced a limit on the spin-dependent WIMP-proton interaction that is a factor of 2 stronger than previously published by the DRIFT collaboration, with a similar exposure.
It is also three orders of magnitude ahead of other directional detectors. The increase in efficiency over our 2012 publication was due to the ability to measure the $z$-coordinate of events, and therefore reject all known DRIFT-IId backgrounds. Further increases in efficiency have already been achieved, and other improvements are in development, including reduced trigger threshold. These results expand the reach of directional dark matter detectors.

\section*{Acknowledgements}
We acknowledge support from the STFC under grant ST/K001337/1, and from the NSF under grant numbers 1103420 and 1103511. JBRB acknowledges support of the Sloan Foundation and the Research Corporation for Science Advancement, and SWS is grateful for STFC grant no. ST/F007337/1. We would also like to thank the owners of Boulby mine, Cleveland Potash Ltd., for their continued support of the underground laboratory.

\section*{Note added in proof}
After this paper was submitted for publication, the NEWAGE directional dark matter experiment published a spin-dependent WIMP-proton cross-section limit from their underground detector (NEWAGE-0.3b') \cite{newage2015}. That limit reaches 557 pb at 200 GeV/c$^2$ WIMP mass. 

\section*{References}

%%%%%%%%%%%%%%%%%%%%%%%%%%%%%%%%%%%%%%%%%%
\begin{comment}
\section{Bibliography styles}

\end{comment}
%%%%%%%%%%%%%%%%%%%%%%%%%%%%%%%%%%%%%%%%%%

\bibliography{MinCarBib}
\end{document}